# Increased perpendicular magnetocrystalline anisotropy governed by magnetic boundary in an asymmetrically terminated FeRh(001) thin film

Eunsung Jekal*

*Department of Materials, ETH Zurich, 8093 Zurich, Switzerland*

Rh-terminated FeRh(001) film is known to be stable in a ferromagnetic (FM) state different from a G-type antiferromagnetic (G-AFM) bulk ground state, while an Fe-terminated FeRh(001) film has the same ground state as the bulk. In this paper, we investigate the magnetic properties of asymmetrically terminated FeRh(001) films: one surface is Fe-terminated and the other is Rh-terminated. Rh surface only ] (RhSO) FM state in asymmetrically terminated FeRh(001) film is identified to exhibit 40% increased perpendicular MCA energy as compare to that of the whole-layer (WL) FM state. This increased MCA energy is governed by Rh atom which is placed at the magnetic boundary. Since FM and G-AFM states are mixed up at the magnetic boundary, $<x2−y2\uparrow|Lz|\ xy\uparrow>$, $<yz,zx\uparrow|Lx|x2−y2\downarrow>$ and $<yz,zx\uparrow|Lx|z2\downarrow>$ couplings which give positive contribution to perpendicular MCA are revealed.

## I. INTRODUCTION

Antiferromangetic (AFM) materials are insensitive to disturbing magnetic fields due to their zero net magnetic moment. This is crucial property to store the information densely and keep it safely in the memory device such as spintronics [1–3]. In addition, the intrinsic high frequencies of AFM dynamics is another benefit of AFM materials over ferromagnetic (FM) ones [4,5]. However, a difficulty in controlling their spin direction is still remained as a major obstacle of AFM material. To overcome this problem, a relativistic spin-orbit coupling-induced torque has been suggested. Another suggestion to manipulating the antiferromagnetic spin has been demonstrated in FeRh, whereby heating and cooling the material can control spin orderings. Moreover, since FeRh has AFM spin state and an antiferromagnetic anisotropic magnetoresistance (AMR) simultaneously, it could be used as a valuable memory device [6–8].

B2 structured FeRh exhibits AFM at room temperature and undergoes the magnetic transition to the FM with volume expansion in just above room temperature 350 K [7]. Since this magnetic phase change with respect to the temperature was discovered over 70 years ago, much theoretical and experimental works have been done in FeRh. Recently, first principle study clarified that bulk and the Fe-terminated FeRh(001) films are ordered as G-type AFM (G-AFM), however, the Rh-terminated films are energetically more stable in FM up to 17 monolayers (ML) [7]. Due to the different magnetic states which strongly depend on surface termination of the film, an asymmetrically terminated film that one surface is Fe and the other surface is Rh terminated has a complicated spin configuration. In films thicker than 8-ML, layers are magnetically distinguished, whereby 3-ML from Rh surface and other rest layers are ordered as FM and G-AFM, respectively. This mixed magnetic state has been defined as Rh surface only (RhSO) FM state in previous work.

In this paper, MCA energy of asymmetrically terminated FeRh(001) film is investigated. In particular, we found that magnetic boundary placed in the asymmetrically terminated film gives increased perpendicular MCA energy as compare to the identical film without magnetic boundary. It can satisfy high bit density and thermal stability, simultaneously, in magnetic data storage.

## II. COMPUTATIONAL METHOD

Vienna ab initio Simulation Package (VASP) [9] is used to perform first-principles calculations. Generalized gradient approximation [10] is employed for the exchange correlation potential. To optimize the structure of the film, internal coordination is fully relaxed while the two dimensional lattice constant was fixed as 2.998 A which is corresponding to G-AFM bulk value. MCA energy is determined by total energy difference between in-plane and perpendicular magnetizations, $E_{MCA} \equiv E(\rightarrow)-E(\uparrow)$.

To obtain reliable MCA energy, 24x24x1 is used for k-mesh in MCA energy calculations while reduced value of 17x17x1 is taken in self-consistent calculations. For wave function expansion, a plane-wave basis set with a cutoff energy of 450 eV was used. To describe the G-AFM spin ordering in the film, we adopt an expanded unit cell of c(2x2) along the xy plane. Film thickness is considered from 4- to 14-ML. Unlike the film calculation, 2x2x2 supercell is applied to obtain MCA energy in the bulk. Since MCA value is very sensitive to k-point density, completely symmetric unit cell has been necessary for calculation accuracy.

## III. RESULTS AND DISCUSSION

The spin configuration of an asymmetrically terminated FeRh(001) thin film is depicted in Fig.1(a).

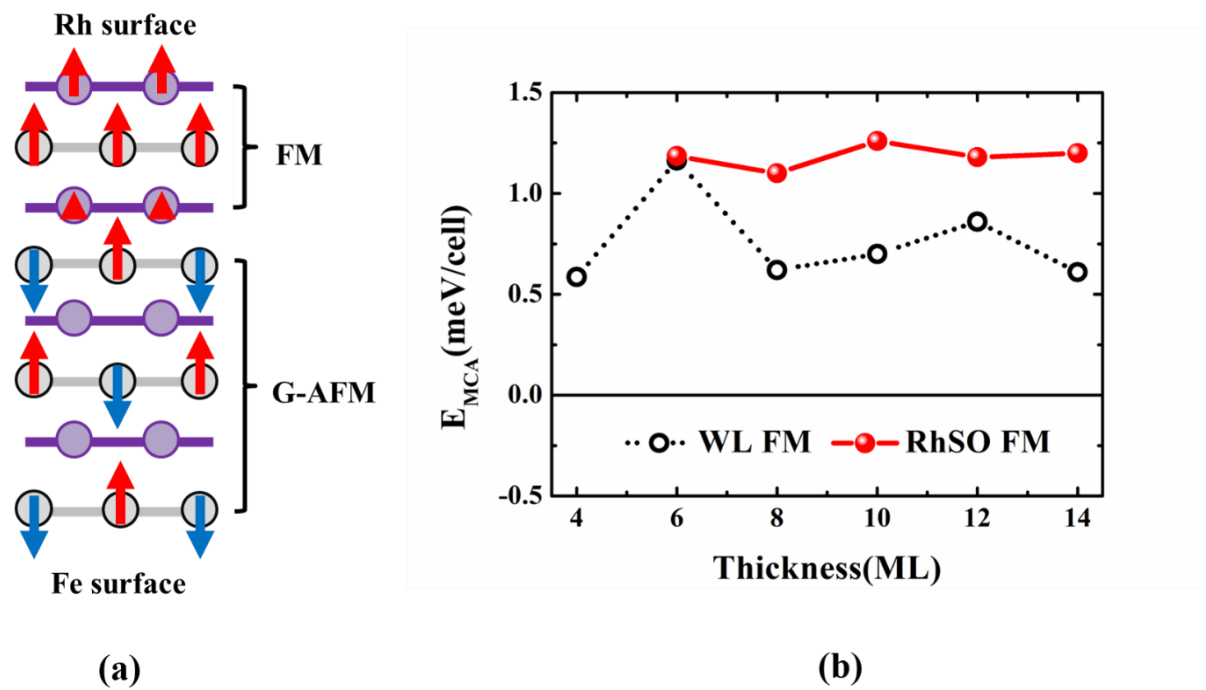

Fig. 1. (a) spin configuration of 8-ML asymmetrically terminated FeRh(001) film. Purple (gray) circle corresponds to Rh (Fe) atom. Red (blue) arrow represents up (down) spin. (b) MCA energy as function of film thickness. Red solid and black dotted lines express the values of RhSO FM and WL FM states, respectively

|  | RhSO FM | WL FM | WK G-AFM |
|---|---|---|---|
| Rh(S) | 1.143 | 1.103 | 0.000 |
| Fe($S_{Rh}$-1) | 3,139(3.142) | 3.162 | ±3.150 |
| Rh($S_{Rh}$-2) | 0.870 | 1.048. | 0.000 |
| Fe($S_{Rh}$-3) | 3.119(-3.132) | 3.127 | ±3.076 |
| Rh($S_{Fe}$-3) | 0.074 | 0.923 | 0.000 |
| Fe($S_{Fe}$-2) | 3.064(-3.075) | 3.164 | ±3.113 |
| Rh(S$_{Fe}$-1) | 0.001 | 0,885 | 0.000 |
| Fe(S) | 3.139(-3.143) | 3.162 | ±3.113 |

Purple and gray circles correspond to Rh and Fe atoms, respectively. To identify the most stable magnetic structure with respect to the film thickness, different number of FM layers from Rh surface is taken into account while other rest layers are fixed as G-AFM [7].

It was clarified that 4- and 6-ML films are stable when whole layers (WL) are ordered as FM, however, the RhSO FM phase as shown in Fig. 1(a) turns out to be the most energetically favorable in thicker than 8-ML films. One crucial point is number of FM layers independents on film thickness except 4- and 6-ML films. The reason of certain FM layers (3-ML from Rh surface) of RhSO FM state has been reasonably explained by magnitude of induced magnetic moment of Rh atoms. Magnetic moments of 8-ML film in RhSO FM, WL FM and WL G-AFM states are listed in Table1, where surface and n layers below Rh (Fe) surface are denoted by (S) and (S$_{Rh(Fe)}$-n), respectively.

Since there are two distinct Fe atoms due to complicated spin interaction at the magnetic boundary in RhSO FM, we denote two different Fe moments. Rh(S) in RhSOFM state has larger magnetic moment by 0.04µB than that of WL FM state while moment of Rh(S$_{Rh}$-2) is much reduced in RhSO FM with respect to the WL FM. This tendency that surface atom of RhSO FM state has large moment but decrease more sharply as compare to WL FM case when n becomes 2 also appears in G-AFM ordered Fe-terminated part. Fe(S) has 0.03µB larger moment in RhSO FM state than WL G-AFM, but in case of Fe(S$_{Fe}$-2), RhSO FM shows smaller value by 0.10µB than WL G-AFM. Interestingly, in all cases, Fe(S$_{Fe}$-1) shows comparable magnetic moments to

Fe(S).

MCA energies of WL FM and RhSO FM states are plotted in Fig.1(b) as function of the film thickness. Dotted and solid lines represent WL FM and RhSO FM, respectively. From the definition of MCA energy in this paper, positive (negative) sign corresponds to perpendicular (in-plane) MCA. Even though perpendicular MCA energy is obtained in both magnetic states, RhSO FM has averagely 40% strong MCA energy than WL FM ones.

While the saturation behavior is clearly shown as the film is getting thicker, the saturated values of WL FM and RhSO FM are 0.70 and 1.25 meV/cell, respectively.

To elucidate the origin of the increased perpendicular MCA of RhSO FM state with respect to the FM state, DOS of WL FM and RhSO FM are compared in Fig. 2

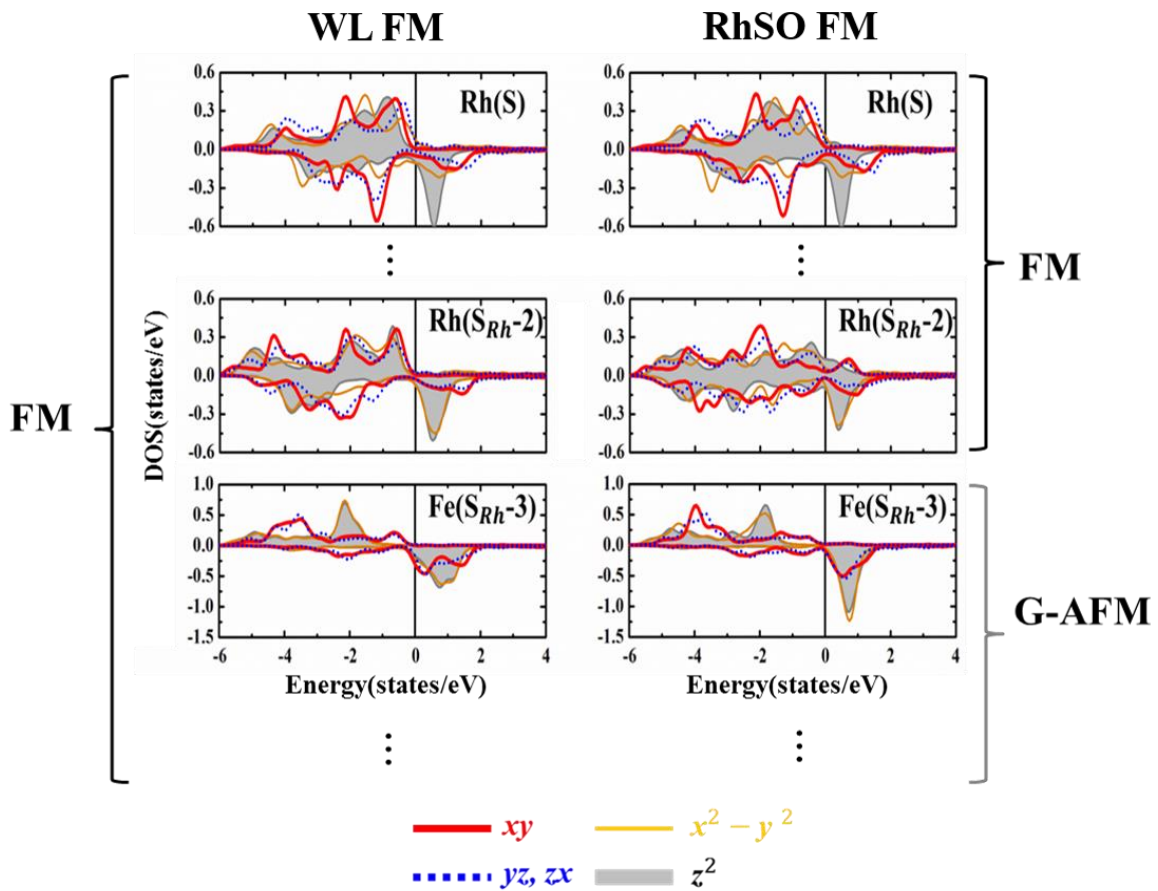

The FM and the G-AFM regions are separated by using black and gray blankets. To emphasize and simplify crucial parts, only Rh(S), Rh($S_{Rh}$-2) and Fe($S_{Rh}$-3) are presented while other layers are expressed as dots(⋯). Because it has been already clarified that most positive contribution to perpendicular MCA is originated by FM ordered Rh(S), and contributions from centered layer are negligible.

For the MCA analysis through DOS, second perturbation theory is adapted [11]. MCA energy is described as

$$E_{MCA}^{\uparrow\uparrow(\downarrow\downarrow)} \approx \xi^2 \sum_{o\uparrow(\downarrow),u\uparrow(\downarrow)} \frac{|<o^{\uparrow(\downarrow)}|L_z|u^{\uparrow(\downarrow)}>|^2 - |<o^{\uparrow(\downarrow)}|L_x|u^{\uparrow(\downarrow)}>|}{\varepsilon_{o\uparrow(\downarrow)} - \varepsilon_{u\uparrow(\downarrow)}}$$

$$E_{MCA}^{\uparrow\downarrow} \approx \xi^2 \sum_{o\uparrow(\downarrow),u\uparrow(\downarrow)} \frac{|<o^{\uparrow(\downarrow)}|L_x|u^{\uparrow(\downarrow)}>|^2 - |<o^{\uparrow(\downarrow)}|L_z|u^{\uparrow(\downarrow)}>|^2}{\varepsilon_{o\uparrow(\downarrow)} - \varepsilon_{u\uparrow(\downarrow)}},$$

where ξ is the spin-orbit coupling constant and ↑(↓) represents majority (minority) spin. o(u) and $\varepsilon_{o(u)}$ represent eigenstate and eigenvalue of occupied(unoccupied) state, respectively.

DOSs of Rh(S) in two magnetic phases look so similar and there is strong <yz,zx↑|$L_x$|$z^2$↓> coupling which gives positive contribution to perpendicular MCA. On the other hands, the DOS shape of magnetic boundary Rh is very different between FM and RhSO FM, in particular near Fermi level. In case of the magnetic boundary Rh with RhSO FM spin configuration, positive contributions from <$x^2-y^2$↑|$L_z$|xy↑>, <yz,zx↑|$L_x$|$x^2-y^2$↓> and < yz,zx↑|$L_x$|$z^2$↓> couplings are much dominant even though negative contribution from < yz,zx↑|$L_x$|xy↑> coupling somewhat reduces perpendicular MCA. Therefore, a net contribution is positive. However, in case of Rh($S_{Rh}$-2) in WL FM state which is corresponding to magnetic boundary Rh of RhSO FM state, there is strong negative contribution, <xy↑|$L_z$|$x^2-y^2$↓>. This coupling is invisible in DOS of magnetic boundary Rh of RhSO FM state.

As a result of DOS analysis, we can argue that increased perpendicular MCA of RhSO FM state is originated by magnetic boundary Rh atom while the strong perpendicular MCA in both magnetic states are mainly determined by FM ordered Rh(S) [15].

For a double check, MCA energies are investigated in bulk with different spin configurations. To design a magnetic boundary in the bulk, one layer and another alternative layer are fixed as G-AFM and FM orderings, respectively. We named this artificial magnetic state as FM+G-AFM.

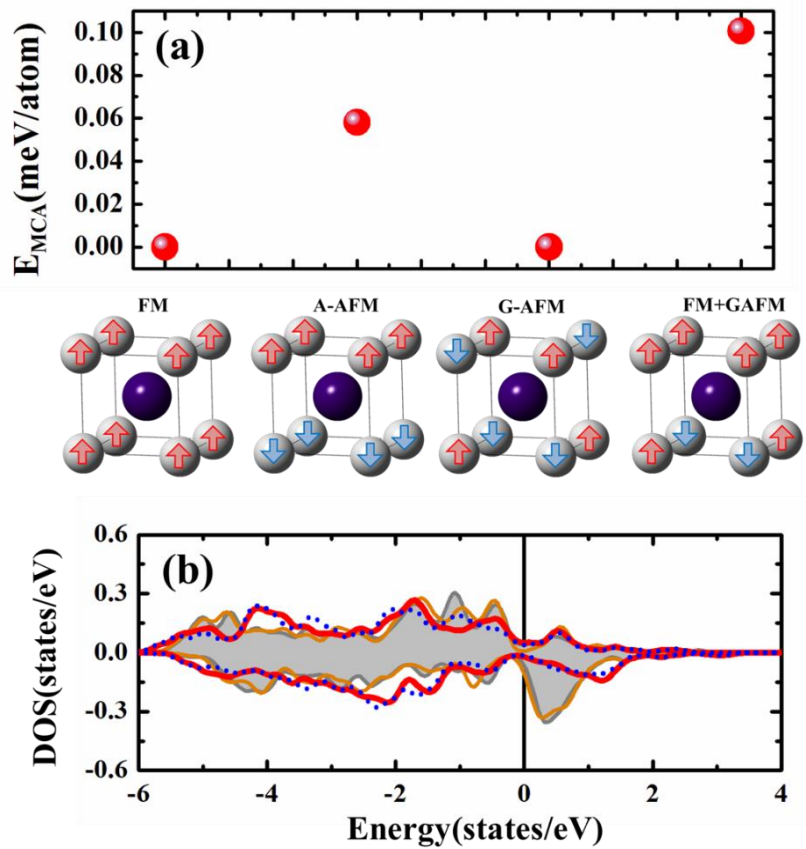

As shown in Fig. 3(a), FM and G-AFM states have zero MCA energy due to a cubic symmetry. Even though A-AFM shows perpendicular MCA energy about 0.06 meV/atom, FM+G-AFM has much stronger MCA energy by 0.10 meV/atom than that of A-AFM. DOS of Rh atom of FM+G-AFM state is presented in Fig. 3(b).

Similar with magnetic boundary Rh in the film (see Fig. 2), strong $<x^2-y^2\uparrow|L_z|xy\uparrow>$ is found even in the bulk. Although a peak of yz,zx↑ orbital which makes <yz,zx↑|$L_x$|$x^2-y^2$↓> and <yz,zx↑|$L_x$|$z^2$↓> couplings near Fermi level is much reduced in the bulk comparison with film case, it still contributes to the perpendicular MCA energy. Therefore, we can confirm the increased perpendicular MCA energy of RhSO FM state is originated by somehow mixed magnetic interactions at the magnetic boundary.

IV. CONCLUSION

In summary, RhSO FM state in asymmetrically terminated FeRh(001) film is identified to exhibit 40% increased perpendicular MCA energy as compare to that of the WL FM state. This increased MCA energy is governed by Rh atom which is placed at the magnetic boundary. Since FM and G-AFM states are mixed up at the magnetic boundary, <$x^2-y^2$↑|$L_z$| xy↑>, <yz,zx↑|$L_x$|$x^2-y^2$↓> and <yz,zx↑|$L_x$|$z^2$↓> couplings which give positive contribution to perpendicular MCA are revealed.